\documentclass[12pt]{article}

\usepackage[dvips]{graphicx}
\usepackage{epsfig}
\usepackage{amsmath,amsfonts,amssymb,amsthm}
\usepackage{verbatim}
\usepackage{psfrag}
\usepackage{bm}
\usepackage{bbm}
\usepackage[square,comma,sort&compress,numbers]{natbib}
\usepackage{color}
\usepackage{slashed}
\usepackage{axodraw4j}

\usepackage{epsf,epsfig}
\usepackage{graphics}

\newcommand{\beq}{\begin{equation}}
\newcommand{\eeq}{\end{equation}}
\newcommand{\bea}{\begin{eqnarray}}
\newcommand{\eea}{\end{eqnarray}}
\newcommand{\vc}[1]{{\textbf{#1}}}

\newcommand{\gsim}
{\;\raisebox{-.3em}{$\stackrel{\displaystyle >}{\sim}$}\;}

\begin{document}
\thispagestyle{empty}

\begin{flushright}
{
\small
TUM-HEP-908-13\\
}
\end{flushright}

\vspace{0.4cm}
\begin{center}
\Large\bf\boldmath
Infrared Correlations in de~Sitter Space:\\
Field Theoretic vs. Stochastic Approach
\unboldmath
\end{center}

\vspace{0.4cm}

\begin{center}
{Bj\"orn~Garbrecht$^a$, Gerasimos Rigopoulos$^{a,b}$ and Yi Zhu$^a$}\\
\vskip0.3cm
{\it $^a$Physik Department T70, James-Franck-Stra{\ss}e,\\
Technische Universit\"at M\"unchen, 85748 Garching, Germany\\
\vskip0.1cm
$^b$Institut f\"ur Theoretische Physik, Philosophenweg 16,\\
Universit\"at Heidelberg, 69120 Heidelberg, Germany}\\
\vskip.3cm
\end{center}

\begin{abstract}
We consider massive $\lambda\phi^4$ theory in de~Sitter background. The mass
of the scalar field $\phi$ is chosen small enough, such that
the amplification of superhorizon momentum modes leads to a significant
enhancement of infrared correlations, but large enough such that perturbation
theory remains valid. Using the Closed-Time-Path approach, we calculate
the infrared corrections to the two-point function of $\phi$ to 2-loop order.
To this approximation, we find agreement with the correlation found using
stochastic methods. When breaking the results down to individual
Feynman diagrams obtained by the two different methods, we observe that
these agree as well.
\end{abstract}


\section{Introduction}

For a free scalar field, that couples to gravitation minimally, there exists no
de~Sitter invariant vacuum state for which the propagator exhibits the light-cone
singularities that are required for a physical massless
field~\cite{Allen:1985ux,Allen:1987tz}.
This is because of the amplification of 
momentum modes that exit the de Sitter horizon.
Due to the redshift, soft modes accumulate on superhorizon scales formally resulting in an infrared (IR) divergence of the propagator,
see {\it e.g.} Ref.~\cite{Tsamis:1993ub}. While it is not clear whether massless
scalar particles are realized in Nature, they can serve as a toy model for gravitons, the propagator of which exhibits similar IR divergences~\cite{Tsamis:1993ub}.

The absence of a de~Sitter invariant vacuum for a massless, minimally coupled,
free scalar field is however physically irrelevant, as there are no interactions which can be used to probe its quantum state. A more interesting and challenging question is whether there is a de~Sitter invariant quantum state for an \emph{interacting} scalar field, and it is this question that has drawn the interest of a number of authors who have addressed the problem using a wide range of methods~\cite{Onemli:2002hr,Onemli:2004mb,Brunier:2004sb,Kahya:2006hc,Riotto:2008mv,Finelli:2008zg,
Janssen:2009pb,Burgess:2009bs,Finelli:2010sh,Burgess:2010dd,Rajaraman:2010xd,Garbrecht:2011gu,Hollands:2011we,
Serreau:2011fu,Jatkar:2011ju,Arai:2011dd,Prokopec:2011ms,Xue:2012wi,Akhmedov:2012pa,
Boyanovsky:2012qs,Arai:2012sh,Boyanovsky:2012nd,Akhmedov:2012dn,Beneke:2012kn,
Serreau:2013psa,Serreau:2013koa,Lazzari:2013boa,Gautier:2013aoa,Tanaka:2013caa,
Lello:2013mfa,Nacir:2013xca,Akhmedov:2013vka}.

The model that has been most widely studied in this respect is $\phi^4$ theory, which is
specified by the Lagrangian
\begin{align}
\label{Lagrangian}
{\cal L}=\sqrt{-g}\left[\frac12
g^{\mu\nu}(\partial_\mu \phi)(\partial_\nu \phi)
-V(\phi)
\right]\,,
\end{align}
where the potential is
\begin{align}
\label{potential}
V(\phi)=\frac 12 m^2 \phi^2
+\frac{\lambda}{4!}\phi^4\,.
\end{align}
The field $\phi$ couples to the de~Sitter background
through the metric tensor $g^{\mu\nu}$. De~Sitter space is parametrised
by the Hubble expansion rate $H$, and various possible choices for
the coordinates are presented
in Ref~\cite{Spradlin:2001pw}. For the present work, we find it useful
to use conformal coordinates on the expanding patch of de~Sitter space
that are given in Eq.~(\ref{dS:metric}) below.

While $m=0$ in the aforementioned massless $\phi^4$ theory, we will take here a non-vanishing mass. The model we consider thus relies on three parameters: $H$, $\lambda$
and $m$. The reason for introducing the mass $m$ is that this
parameter can force the perturbative loop expansion to be valid.
In Minkowski space, perturbation expansion can be performed
provided $|\lambda|<4\pi$. However, in de~Sitter space and when
$m\ll H$, this is no longer true
due to the enhancement of IR modes of the scalar field $\phi$.
As we discuss below, the parametric region where there is a significant
IR enhancement of superhorizon modes but, at the same time, perturbation theory
remains valid, is given by
\begin{align}
\label{m:lambda:relations:IR:pert}
m^2\ll H^2\quad\textnormal{and}\quad\lambda\ll m^4/H^4\,.
\end{align}
We refer to the model in this parametric domain as
light, perturbative $\phi^4$ theory in de~Sitter space.

\begin{figure}[t!]
\begin{center}
\parbox{3.1cm}
{
\center
\vskip1.25cm
\epsfig{file=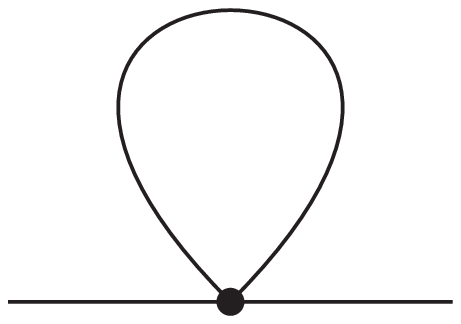,scale=0.6}

(A)
}
\parbox{5.1cm}
{
\center
\vskip1.25cm
\epsfig{file=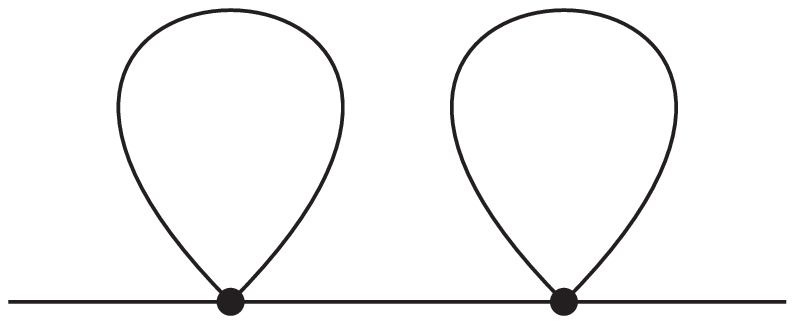,scale=0.6}

(B)
}
\parbox{3.1cm}
{
\center
\epsfig{file=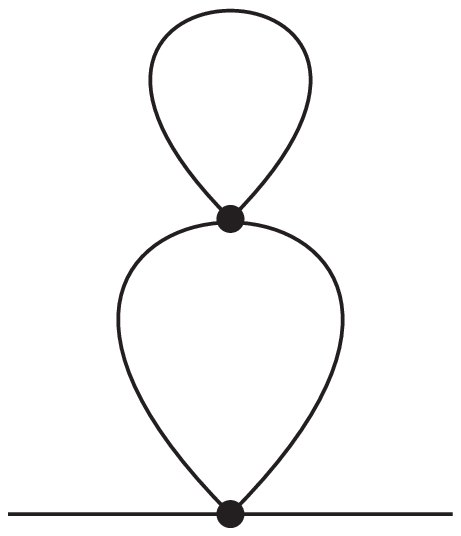,scale=0.6}

(C)
}
\parbox{3.1cm}
{
\center
\vskip1.40cm
\epsfig{file=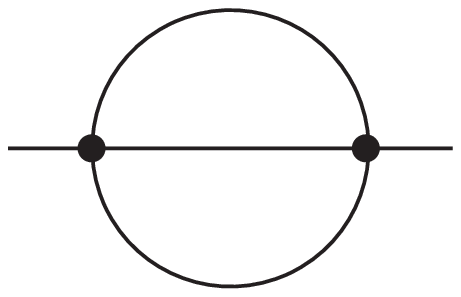,scale=0.6}

(D)
}
\end{center}
\caption{
\label{fig:Feyn:diagrams}
The diagrammatic contributions to the self energy up to order $\lambda^2$.
When amputating the external lines, Diagram~(A) corresponds to the
seagull-type self-energy ${\rm i}\Pi_{\rm sg}$, Diagram~(C)
to the cactus-type self-energy ${\rm i}\Pi_{\rm ca}$ and Diagram~(D)
to the sunset-type self-energy ${\rm i}\Pi_{\rm ss}$.
Accordingly, for the diagrammatic decomposition within the stochastic approach,
we denote Diagram~(A) by $\langle \phi^2\rangle_{\rm sg}$,
Diagram~(B) by$\langle \phi^2\rangle_{\rm sgsg}$,
Diagram~(C) by$\langle \phi^2\rangle_{\rm ca}$ and
Diagram~(D) by$\langle \phi^2\rangle_{\rm ss}$.
}
\end{figure}

The quantity that we aim to calculate is the fluctuation of $\phi$,
{\it i.e.} the expectation value
$\langle \phi^2\rangle$. We suppress here the space-time coordinates of the
field operators, but imply that the separation of the two coordinates should
be taken to be of superhorizon scale, $\gsim H^{-1}$, what we specify more precisely
in the calculations below. In Section~\ref{sec:QFT}, we
pursue the direct approach to this
problem, which is to use Quantum Field Theory (QFT) and to
calculate the loop corrections to the the propagator. The background of
a curved space-time suggests to use the Closed-Time-Path (CTP) formalism for this
purpose.
Such a calculation must in particular address two points:
\begin{itemize}
\item
In the massless $\phi^4$ model, the loop expansion appears not to converge,
as can be seen from the corresponding problem in Euclidean de~Sitter space and as it
is also indicated for Lorentzian de~Sitter space in the present work. Promoting the
scalar field to an $O(N)$ multiplet, a $1/N$ expansion can be performed and
be truncated at the zeroth order, which includes only the one-loop
seagull diagram in Figure~\ref{fig:Feyn:diagrams}(A). This calculation
has been performed in Ref.~\cite{Riotto:2008mv} and confirmed
in some subsequent studies~\cite{Garbrecht:2011gu,Serreau:2011fu,Prokopec:2011ms}. To our knowledge,
an extension to order $1/N$ or beyond has not yet been performed.
As stated above, in the present work we force the convergence of the loop
expansion through a non-vanishing mass term that
satisfies the condition~(\ref{m:lambda:relations:IR:pert}).
\item
The leading order correction in both the $O(N)$ symmetric model
as well as in the light, perturbative $\phi^4$ model is given by the
seagull diagram [Figure~\ref{fig:Feyn:diagrams}(A)].
It plays a special role, because it is a local correction and
therefore can be absorbed in the redefinition of the local mass term.
At higher orders, there occur non-local diagrams as well, and there
is no agreement in the current literature about how to correctly evaluate these.
In the light massive model we need to evaluate the sunset diagram
[Figure~\ref{fig:Feyn:diagrams}(D)] for a consistent calculation of
the fluctuation of $\phi$ to ${\cal O}(\lambda^2)$. Here, we do so by
evaluating, in a rather straightforward manner, the convolution integrals corresponding to the sunset diagram, which appear
in the Schwinger-Dyson equation. For this
procedure, it turns out to be crucial to account for the decay of the IR fluctuations
of the scalar field at very large distances. The computation of the
${\cal O}(\lambda^2)$ correction to the fluctuation of $\phi$ is therefore the
main technical result presented here.
\end{itemize}

It is desirable to perform a consistency check of the QFT
calculation. Substantial progress has been made for massless $\phi^4$ theory in
Euclidean de~Sitter space, where an invariant quantum state is derived to
leading IR order in Ref.~\cite{Rajaraman:2010xd}.
This calculation is confirmed in Ref.~\cite{Beneke:2012kn},
where a loop expansion and the necessary resummation are performed. Moreover,
it is pointed out there that the Euclidean QFT result agrees
with what is obtained from the stochastic approach for scalar field fluctuations in
Lorentzian de~Sitter space~\cite{Starobinsky:1986fx,Starobinsky:1994bd}.
Besides, there is work arguing that in de~Sitter space
Euclidean two-point functions can be analytically continued in order to obtain their Lorentzian counterparts~\cite{Higuchi:2010xt}.
Therefore, it appears interesting to compare the
QFT result with the stochastic result. For this purpose, we formulate in
Section~\ref{sec:stochastic} the stochastic approach in terms of a diagrammatic expansion that bears a close relation to the CTP diagrams and indeed find agreement to order ${\cal O}(\lambda^2)$. In fact, the agreement extends to a diagram-by-diagram comparison between the two approaches.

We emphasise here that both the QFT calculation (using the CTP formalism)
and the stochastic approach are based on the field quantisation of $\phi$.
The primary purpose of the nomenclature ``QFT'' and ``stochastic'' is therefore
to distinguish between these two methods. However, it also reflects the fact that in the QFT approach no assumption about the classical behaviour
of the IR modes is made, in contrast to the stochastic method.

A setup similar to the one specified
by the Lagrangian~(\ref{Lagrangian})
and the relations~(\ref{m:lambda:relations:IR:pert})
is studied within Ref.~\cite{Gautier:2013aoa}.
Although the methodology agrees to some extent with
what is used here in that the CTP formalism is employed, there are differences in the details of the calculation and in the choice of
the quantities that are presented as the final results.
A perturbative calculation in massless $\phi^4$ theory is also valid at early
times, when using a de~Sitter breaking propagator with IR correlations,
that grow in time, see {\it e.g.} Refs.~\cite{Onemli:2002hr,Onemli:2004mb,Brunier:2004sb,Kahya:2006hc}.
In this setup, the agreement
between stochastic and QFT results at order $\lambda$ has been noted
in Ref.~\cite{Kahya:2006hc}. For other earlier discussions of the stochastic-QFT correspondence see e.g \cite{Tsamis:2005hd, Finelli:2008zg}.

For most of the present discussion, we work in four space-time dimensions.
This way, we avoid the recurring notation of factors, that account for a general
dimensionality. The generalisation to $D$ space-time dimensions is however
straightforward, and a brief discussion along with the main results is presented
in Section~\ref{sec:Ddim}.

\section{Stochastic Approach}
\label{sec:stochastic}

In the stochastic approach~\cite{Starobinsky:1986fx},
the field $\phi$ is separated
into a long wavelength ({\it i.e.} superhorizon) part,
that is treated as a classical
stochastic variable, and a short wavelength part for which the
underlying description as a quantum field is maintained. 
For simplicity \emph{we here denote the long wavelength part by $\phi$ when
we refer to the stochastic approach}, using the same
symbol as for the underlying field. When it is assumed
that the behaviour of the long wavelength modes is classical,
their dynamics is driven by a stochastic noise induced by the quantum short-wavelength modes. In particular the stochastic theory of inflationary dynamics is based on the Starobinsky equation
\beq\label{staro1}
\dot{\phi}+\frac{\partial_\phi V}{3H} = \xi(t)\,,
\eeq
a Langevin-type equation for the scalar field where $\xi$ is a Gaussian random force with
\beq
\langle\xi(t)\xi(t')\rangle = \frac{H^3}{4\pi^2}\delta(t-t')\,.
\eeq
The expectation value of an operator $\mathcal{O}[\phi]$ is given by\footnote{Note that we normalise the functional integration measures that
appear in Eq.~(\ref{func:exp:stoch}) such that
$\langle \mathbbm{1}\rangle=1$. This also corresponds to the retarded Ito regularisation of the stochastic equation~(\ref{staro1})~\cite{Kamenev:2009jj}.}
\beq
\label{func:exp:stoch}
\langle \mathcal{O}[\phi] \rangle = \int D\!\left[\xi\right] {\rm e}^{-\frac{1}{2}\int \!\! dt\,\xi^2\frac{4\pi^2}{H^3}}\int D\!\left[\phi\right]\mathcal{O}[\phi]\,\delta\!\left(\dot{\phi}+{\partial_\phi V}/{3H} - \xi\right)\,.
\eeq
By expressing the delta functional as a functional ``Fourier transform'' with the aid of an auxiliary field $\psi$ and performing the Gaussian $\xi$ integral we obtain
\beq\label{GenFunc}
\langle \mathcal{O}[\phi] \rangle=\int D\!\left[\phi, \psi\right] \mathcal{O}[\phi] \,{\rm e}^{-\int \!\!dt \,\left[ {\rm i}\psi\left(\dot{\phi}+\frac{ \partial_\phi V}{3H}\right) + \frac{H^3}{8\pi^2}\psi^2\right]}\,.
\eeq

Let us now focus on the quadratic potential $V(\phi)=\frac{1}{2}m^2\phi^2+\frac{\lambda}{4!}\phi^4$. To obtain a diagrammatic expansion we rewrite the action by bringing the quadratic part in a more symmetric form\footnote{In fact, the analogue of Eq.~(\ref{GenFunc2}) containing second time derivatives can be obtained directly from the more fundamental CTP path integral after short wavelength modes are integrated out~\cite{Rigopoulos:2013exa, GRZ}.}
\beq\label{GenFunc2}
\langle \mathcal{O}[\phi] \rangle=\int D\!\left[\phi, \psi\right] \mathcal{O}[\phi] \,{\rm e}^{-{\rm i}\int \!\!dt \, \left[ \frac{1}{2}\left(\begin{smallmatrix}\phi\,, &\psi \end{smallmatrix}\right)\left(\begin{smallmatrix}0 & -\partial_t+\frac{m^2}{3H}\\ \partial_t +\frac{m^2}{3H} & -{\rm i}\frac{H^3}{4\pi^2}\end{smallmatrix}\right)
\left(\begin{smallmatrix}\phi \\ \psi \end{smallmatrix}\right) + \frac{\lambda}{3!}\frac{\psi\phi^3}{3H}\right]}\,.
\eeq

\begin{figure}[t!]\label{fig:Stoch-propagators}

\begin{center}
\parbox{3.1cm}
{
\center
\vskip1.25cm
\epsfig{file=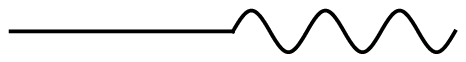,scale=0.6}

$-{\rm i}G^R(t,t')$
}
\parbox{3.1cm}
{
\center
\vskip1.25cm
\epsfig{file=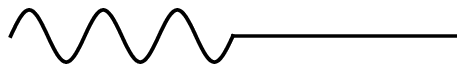,scale=0.6}

$-{\rm i}G^A(t,t')$
}
\parbox{3.1cm}
{
\center
\vskip1.35cm
\epsfig{file=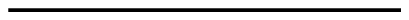,scale=0.6}

\vskip0.1cm
$F(t,t')$
}
\parbox{3.1cm}
{
\center
\vskip1.25cm
\epsfig{file=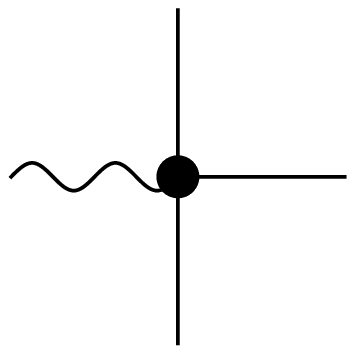,scale=0.6}

$-{\rm i}\frac{\lambda}{3H}\int d\tau$
}
\end{center}
\caption{The elements out of which stochastic diagrams are constructed. The choice of vertex factor implies that the assembled diagrams should be divided by their symmetry factor.
}
\end{figure}

The free correlation functions are then determined as the functional and matrix inverse of the quadratic operator:
\beq\label{propagators}
\left(\begin{smallmatrix}\langle\phi(t)\phi(t')\rangle & \langle\phi(t)\psi(t')\rangle \\ \langle\psi(t)\phi(t')\rangle & \langle\psi(t)\psi(t')\rangle\end{smallmatrix}\right)\equiv-{\rm i}\left(\begin{smallmatrix}0 & \left(-\partial_t+\frac{m^2}{3H}\right)\\ \left(\partial_t +\frac{m^2}{3H}\right) & -{\rm i}\frac{H^3}{4\pi^2}\end{smallmatrix}\right)^{-1}\!\!\delta(t-t')=
\left(\begin{smallmatrix}
F(t,t')&-{\rm i}G^R(t,t')\\
-{\rm i}G^A(t,t')&0\end{smallmatrix}\right).
\eeq
Here $G^{(R,A)}(t,t')$ are the retarded and advanced Green functions for the operator $\partial_t +\frac{m^2}{3H}$
\beq
G^R(t,t')=G^A(t',t)=
{\rm e}^{-\frac{m^2}{3H}(t-t')}\Theta(t-t')\,,
\eeq
and $F(t,t')\equiv\langle\phi(t)\phi(t')\rangle$ is the 2-point function of $\phi$
\beq\label{corr}
F(t,t')= \frac{H^3}{4\pi^2}\int\limits_{0}^{+\infty} d\tau \,\,G^R(t,\tau)G^A(\tau,t')=\frac{3H^4}{8\pi^2m^2}\left({\rm e}^{-\frac{m^2}{3H}|t-t'|}-{\rm e}^{-\frac{m^2}{3H}(t+t')}\right)\,.
\eeq
If $t$ and $t'$ are taken to be sufficiently large, or, equivalently, the stochastic process is taken to have begun early enough, the correlator reduces to
\beq\label{corr2}
F(t,t')\simeq \frac{3H^4}{8\pi^2m^2}{\rm e}^{-\frac{m^2}{3H}|t-t'|}
\eeq
which is the form that we'll be using from now on. Note that in the massless limit, $G^R(t,t')\rightarrow \Theta(t-t')$,
and the variance, as is well known \cite{Vilenkin:1983xp}, grows linearly with time
\beq
\langle\phi^2(t)\rangle_{m=0}\simeq \frac{H^3}{4\pi^2}t.
\eeq

We can now construct a diagrammatic expansion out of the elements shown in Figure~\ref{fig:Stoch-propagators}: there are three types of propagator, $-{\rm i}G^R(t,t')$, $-{\rm i}G^A(t,t')$ and $F(t,t')$ along with the relation $G^R(t,t')=G^A(t',t)$ and a single vertex with one wiggly and three solid legs. A vertex refers to an internal time variable which is integrated over. Note that the diagrammatic elements are identical in form to those of the CTP formalism, expressed in the Keldysh basis, but with the additional three-wiggle-line vertex omitted. The absence of the latter vertex corresponds to the semiclassical nature of the result. In this work we focus on the two-point function to second order, given by the diagrams of Figure~\ref{fig:Stoch-Feyn}. Note that vacuum bubble diagrams are zero. To facilitate later comparison with the QFT calculation, we break down the result into the individual Feynman diagrams depicted in Figure~\ref{fig:Stoch-Feyn}. Setting $t=t'$ and taking the late time limit we obtain:

\begin{figure}[t!]
\parbox{8cm}
{\vskip0.5cm
\epsfig{file=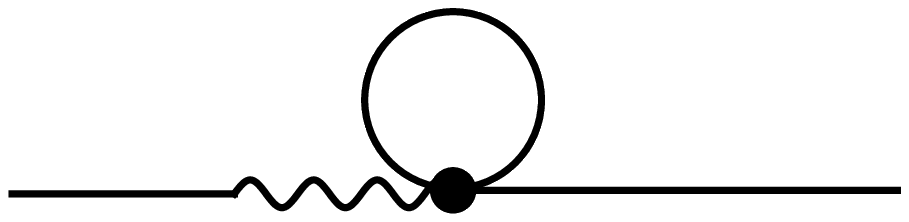,scale=0.36}\hskip0.2cm $+$
\epsfig{file=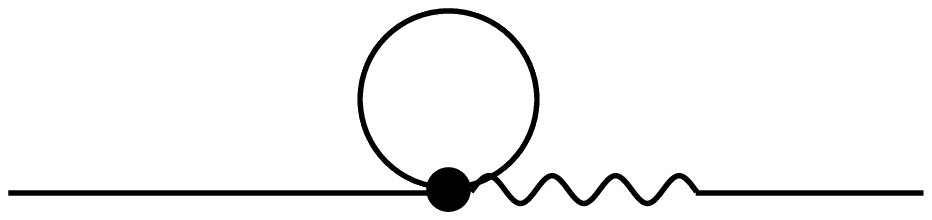,scale=0.36}
\vskip0.3cm
$\langle\phi^2\rangle_{\rm sg}$
}
\parbox{10cm}
{
\epsfig{file=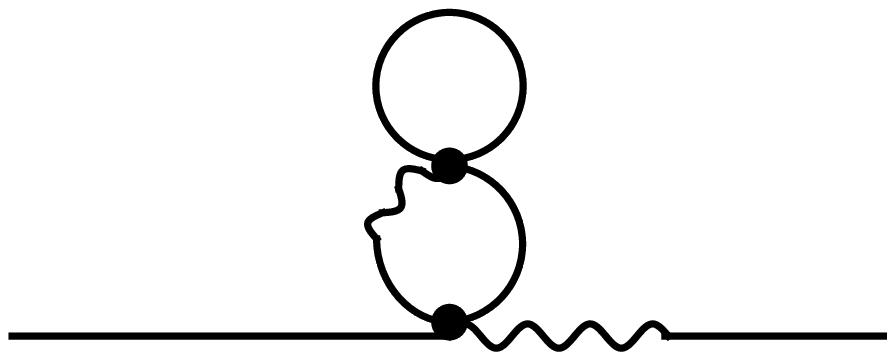,scale=0.36}
$+$
\epsfig{file=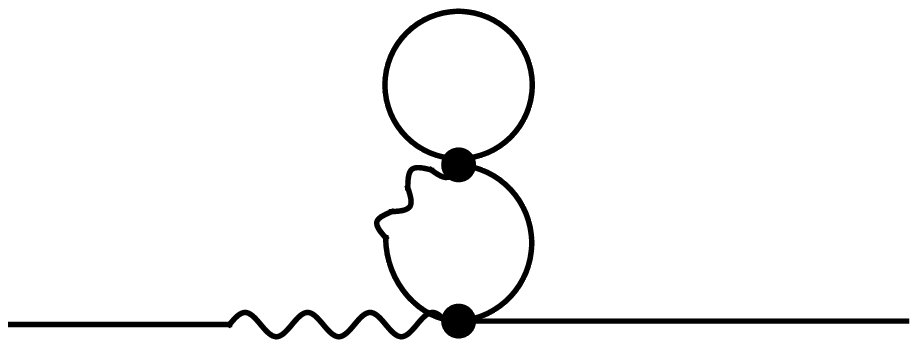,scale=0.36}
\vskip0.3cm
$\langle\phi^2\rangle_{{\rm ca}}$
}
\vskip0.8cm
\parbox{15cm}
{
\epsfig{file=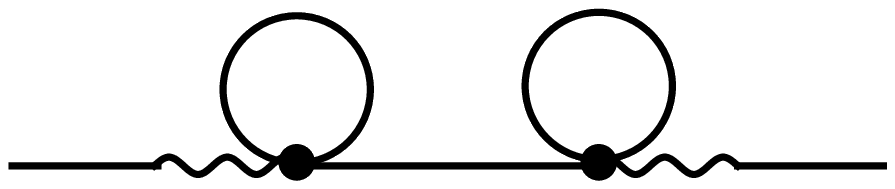,scale=0.36}
$+$
\epsfig{file=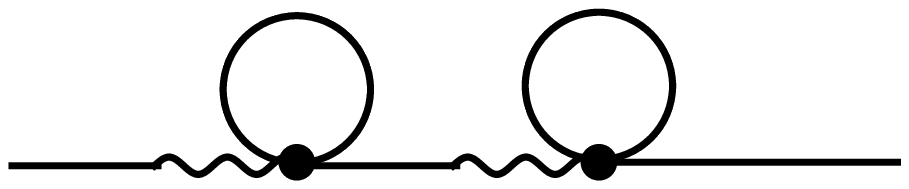,scale=0.36}
$+$
\epsfig{file=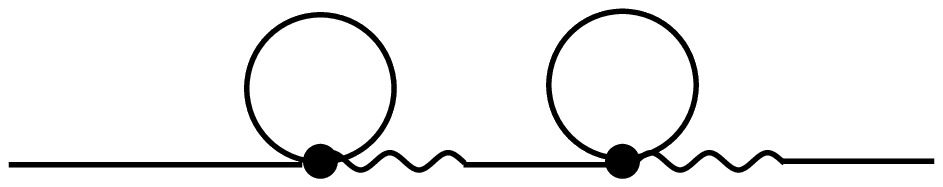,scale=0.36}
\vskip0.3cm
$\langle\phi^2\rangle_{{\rm sgsg}}$
}
\vskip0.8cm
\parbox{15cm}
{
\epsfig{file=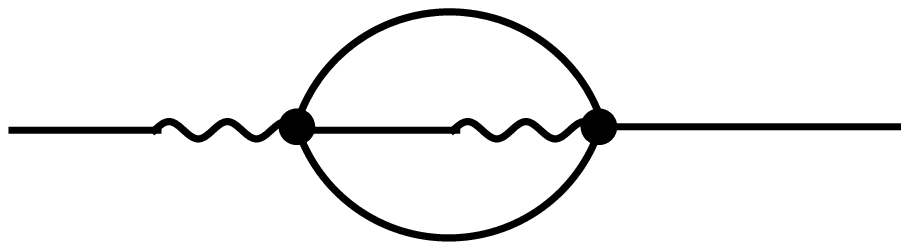,scale=0.36}$+$
\epsfig{file=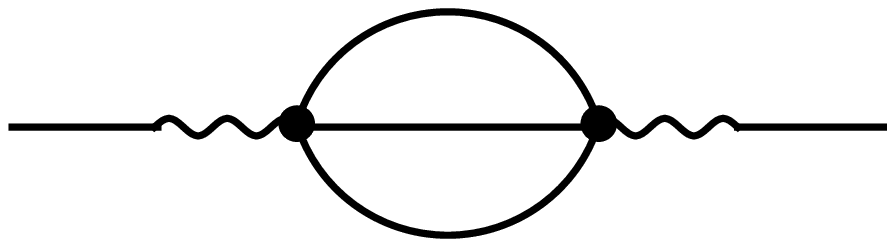,scale=0.36}$+$
\epsfig{file=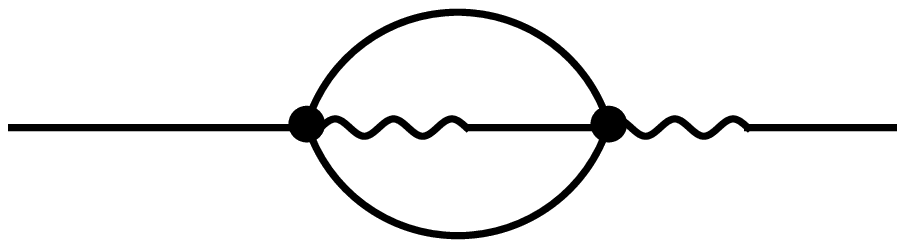,scale=0.36}
\vskip0.3cm
$\langle\phi^2\rangle_{{\rm ss}}$
}
\caption{\label{fig:Stoch-Feyn}
The stochastic diagrams contributing to $\langle\phi(t)\phi(t')\rangle$ up to order $\lambda^2$. Vacuum bubbles are zero by construction. Note that they are identical in form to the CTP diagrams in the Keldysh basis but with the three-wiggle vertex removed. This corresponds to a semiclassical approximation. The labels refer to the topology of the diagrams: seagull, cactus, double seagull and sunset.
}
\end{figure}

\begin{subequations}
\label{stoch:diagrams}
\begin{align}\label{sg}
\langle\phi^2\rangle_{{\rm sg}}=& - \lambda\frac{9H^8}{128\pi^4m^6}\,,
\displaybreak[0]\\
\label{ca}
\langle\phi^2\rangle_{{\rm ca}}=&{\lambda^2}\frac{27H^{12}}{2048\pi^6m^{10}}\,,
\displaybreak[0]\\
\label{sgsg1}
\langle\phi^2\rangle_{{\rm sgsg}}=&{\lambda^2}\frac{27H^{12}}{2048\pi^6m^{10}}\,,
\displaybreak[0]\\
\label{ss1}
\langle\phi^2\rangle_{{\rm ss}}=&{\lambda^2}\frac{ 9H^{12}}{ 1024\pi^6m^{10}}\,.
\end{align}
\end{subequations}
Adding up the individual contributions, we obtain
\beq\label{lambda-series1}
\lim\limits_{t\rightarrow\infty}\langle\phi(t)^2\rangle= \frac{3H^4}{8\pi^2m^2} - \lambda\frac{9H^8}{128\pi^4m^6} + \lambda^2\frac{9H^{12}}{256\pi^6m^{10}}\,.
\eeq
As we will see below, the result (\ref{lambda-series1}) agrees with the result from the QFT Schwinger-Dyson equations. Furthermore, each individual contribution from the topologically distinct diagrams, Eqs.~(\ref{sg})-(\ref{ss1}), equals the QFT contribution from diagrams of corresponding topology.

We should note here a different way in which the result (\ref{lambda-series1}) can be obtained. At late times the stochastic process (\ref{staro1}) is described by the probability distribution
function~\cite{Starobinsky:1986fx,Starobinsky:1994bd}
\begin{align}
\label{stoch:PDF}
\varrho(\phi)={\cal N}{\rm e}^{-\frac{8\pi^2}{3H^4} V(\phi)}\,,
\end{align}
where the normalisation ${\cal N}$ is determined by the condition
\begin{align}
\label{PDF:normalisation}
\int\limits_{-\infty}^\infty d\phi \varrho(\phi)=1\,.
\end{align}
Expectation values (at equal times) are obtained using the probability distribution
function in the usual way, for example
\begin{align}
\label{phi:expectationvalue}
\langle \phi^n \rangle
=\int\limits_{-\infty}^\infty d\phi \phi^n \varrho(\phi)\,.
\end{align}
For the particular potential~(\ref{potential}), we can expand
\begin{align}
{\rm e}^{-\frac{8\pi^2}{3H^4}V(\phi)}
={\rm e}^{-\frac{4\pi^2}{3H^4} m^2 \phi^2}
\left(
1-\frac{\pi^2 \lambda \phi^4}{9H^4}+\frac{\pi^4\lambda^2\phi^8}{162 H^8}+\cdots
\right)\,.
\end{align}
Substituting this into Eq.~(\ref{phi:expectationvalue}), we perform
the integrals for the fluctuation and the normalisation with the result
\begin{align}
\label{fluc:massive}
\langle \phi^2 \rangle=\frac{3H^4}{8\pi^2 m^2}-\frac{9\lambda H^8}{128\pi^4m^6}+\frac{9\lambda^2 H^{12}}{256 \pi^6 m^{10}}+\cdots\,,
\end{align}
which coincides with Eq.~(\ref{lambda-series1}).
Again, the expansion~(\ref{fluc:massive}) has an immediate interpretation in terms
of Feynman diagrams: For each vertex, we assign a factor
$-\lambda 8\pi^2/(3H^4)$ and for each propagator, a factor
$3H^4/(8\pi^2 m^2)$. Moreover, we divide by the
appropriate symmetry factor.
We thus obtain
for the seagull diagram [Figure~\ref{fig:Feyn:diagrams}(A), symmetry factor $2$]
$\langle\phi^2\rangle_{{\rm sg}}$,
for the diagram in Figure~\ref{fig:Feyn:diagrams}(B) (symmetry factor $4$)
$\langle\phi^2\rangle_{{\rm sgsg}}$, for the
cactus diagram [Figure~\ref{fig:Feyn:diagrams}(C), symmetry factor $4$] $\langle\phi^2\rangle_{{\rm ca}}$,
and for the sunset diagram
[Figure~\ref{fig:Feyn:diagrams}(C), symmetry factor $6$] $\langle\phi^2\rangle_{{\rm ss}}$, where all results can be found in Eqs.~(\ref{stoch:diagrams}).

Apparently, the expansion (\ref{lambda-series1}) is valid provided
$\lambda\ll m^4/H^4$, in accordance with relation~(\ref{m:lambda:relations:IR:pert}). However, the stochastic theory implies that the regime with $\lambda > m^4/H^4$  is also meaningful since the series can be summed and correlation functions for the
potential~(\ref{potential}) can be evaluated exactly using Eq~(\ref{stoch:PDF}). In this case the integrals leading to $\langle \phi^2\rangle$ can be evaluated in terms of Bessel and of Hypergeometric functions.

Before closing this Section on the stochastic approach to inflation, we discuss the spatial correlations of the stochastic field.
We use conformal coordinates with the metric tensor
\begin{align}
\label{dS:metric}
g_{\mu\nu}(x)=a^2(\eta){\rm diag}(1,-1,-1,-1)\,.
\end{align}
The scale factor is given by $a(\eta)=-1/(H\eta)$, and $\eta=x^0$
is the conformal time.
Below, we sometimes write $a(x)\equiv a(x^0)$ for a 4-vector $x$.
For the expanding de~Sitter space,
$\eta\in (-\infty;0)$. Now we
first consider two points
$\mathbf x$ and $\mathbf x^\prime$
that begin at a time $\eta_0$
with the same field value and that are initially separated by a physical distance
$\Delta r(\eta_0) =a(\eta_0) |\Delta \mathbf x|\sim 1/H$, which is the smallest distance for which the stochastic description is meaningful and where $\Delta \mathbf x=\mathbf x-\mathbf x^\prime$. The subsequent evolution of the field $\phi$
will be given by Eq.~(\ref{staro1}) for two different realisations of the noise field $\xi$. From Eq.~(\ref{corr}) we see that the field values at these two points will be completely uncorrelated after a physical time interval of $\Delta t \sim \frac{H}{m^2}$,
where the physical time $t$ is related to the conformal time through $dt=a(\eta) d\eta$. Here, we fix the physical time coordinate by setting
$a(t)\equiv a(\eta)=a_0{\rm e}^{H t}$, and accordingly for the
time variables with primes or subscripts.
The physical separation of these two points will then be
$\Delta r (\eta)\sim \frac{1}{H} {\rm e}^{\frac{H^2}{m^2}}$,
where $t=t_0+\Delta t$. Thus the field $\phi$ maintains its coherence over distances that satisfy
\beq
1<a^2(\eta)H^2\Delta\vc{x}^2 \ll {\rm e}^{\frac{2H^2}{m^2}}\,,
\eeq
while it is incoherent on physical scales $\Delta r > \frac{1}{H} {\rm e}^{\frac{H^2}{m^2}}$. Beyond this general statement, it is in fact possible to obtain the dependence of the correlation function on spatial separation by invoking de Sitter invariance.
When we write $\Delta x=x-x^\prime$, such that
\begin{align}
\Delta x^2=(\eta-\eta^\prime)^2-(\mathbf x-\mathbf x^\prime)^2\,,
\end{align}
where $\mathbf x=(x^1,x^2,x^3)$,
we can define a de~Sitter
invariant length function as~\cite{Onemli:2002hr,Tsamis:1992xa}
\begin{align}
\label{length:fun}
y(x;x^\prime)=a(\eta)a(\eta^\prime)H^2\Delta x^2=\frac{\Delta x^2}{\eta\eta^\prime}=-4\sin^2\left(\frac 12 H \ell(x;x^\prime)\right)
\,.
\end{align}
We have indicated here the relation to the length $\ell(x;x^\prime)$ along
a geodesic that connects these points.
Note that for $y>0$, $\ell(x;x^\prime)$ is purely imaginary, corresponding
to time-like separations. Space-like separations correspond to $-4<y<0$,
where $\ell$ is real, while for $y<-4$, there is no geodesic that connects
the two points (even though a complex $\ell$ may still be defined),
see {\it e.g.} Ref.~\cite{Tsamis:1992xa}.
Due to de~Sitter invariance, the correlations of $\phi$ should be functions
of $y$ only.
For large physical time separations $t - t^\prime\gg 1/H$, we can approximate
\beq
y(x;x^\prime)\approx {\rm e}^{H(t-t^\prime)}-{\rm e}^{H(t+t^\prime)}a_0^2 H^2|\Delta\vc{x}|^2\,,
\eeq
while for separations with $t=t^\prime$
\begin{align}
y(x;x^\prime)=-{\rm e}^{2Ht}a_0^2 H^2|\Delta\vc{x}|^2\,.
\end{align}
We therefore replace
$e^{-H(t-t')} \rightarrow \frac{1}{a^2(t)H^2|\Delta\vc{x}|^2}$ in
Eq.~(\ref{corr}) to obtain
\beq
\langle\phi(t,\vc{x})\phi(t,\vc{x}')\rangle =\frac{3H^4}{8\pi^2m^2}\left(\frac{1}{a^2(t) H^2|\vc{x}-\vc{x}'|^2}\right)^\frac{m^2}{3H^2}\,.
\eeq
We thus see a mild power-law decay at large distances. This relation is verified in the the following Section, see Eq.~(\ref{prop:free:decay}).
Besides, we have also ignored here the sign of $y$, which will be
properly accounted for in the QFT approach as well.


\section{Field Theory Approach}
\label{sec:QFT}

\subsection{Propagator for a Massive Free Field}

We now pursue the QFT approach to light, perturbative
$\phi^4$ theory in de~Sitter space. Due to the time-dependent background, it is pertinent to use the CTP approach.
As by the relations~(\ref{m:lambda:relations:IR:pert}),
the problem is perturbative and we can pursue a loop expansion
that we truncate here at the two-loop order.
The result will be a perturbatively improved propagator that we can compare
with the fluctuation of $\phi$ that is obtained by stochastic means,
Eq.~(\ref{lambda-series1}).

The basic building block of the
Feynman diagrams is the free propagator ${\rm i}\Delta^{(0)}$,
that satisfies the
Klein-Gordon equation
\begin{align}
\label{KG:free}
a^4\left(\nabla_x^2-m^2_{}\right){\rm i}\Delta^{(0)fg}(x;x^\prime)
=fg \delta^{fg}\,{\rm i}\delta^4(x-x^\prime)
\,,
\end{align}
where $f,g=\pm$ are CTP indices and $(\nabla_x)_\mu$ is the covariant derivative
with respect to $x$.
Note that the IR effects in de~Sitter space
can at least partly be accounted
for by a dynamical mass
$m_{\rm dyn}$~\cite{Garbrecht:2011gu,Beneke:2012kn,Gautier:2013aoa}.
When this is the case, the
leading IR effects are captured by a full propagator ${\rm i}\Delta$
that also satisfies
the Klein-Gordon equation with $m$ replaced by $m_{\rm dyn}$. Therefore,
Eq.~(\ref{KG:free}) and its solution also describe important properties
of the full propagator in the interacting theory.

The causal properties of two-point functions are accounted for
by the following $\varepsilon$-prescriptions:
\begin{subequations}
\label{eps:causal}
\begin{align}
{\Delta x^{T}}^2(x;x^\prime)=
{\Delta x^{++}}^2(x;x^\prime)
=&(|\eta-\eta^\prime|-{\rm i}\varepsilon)^2-|\mathbf x -\mathbf x^\prime|^2\,,\\
{\Delta x^{<}}^2(x;x^\prime)=
{\Delta x^{+-}}^2(x;x^\prime)
=&(\eta-\eta^\prime+{\rm i}\varepsilon)^2-|\mathbf x -\mathbf x^\prime|^2\,,\\
{\Delta x^{>}}^2(x;x^\prime)=
{\Delta x^{-+}}^2(x;x^\prime)
=&(\eta-\eta^\prime-{\rm i}\varepsilon)^2-|\mathbf x -\mathbf x^\prime|^2\,,\\
{\Delta x^{\bar T}}^2(x;x^\prime)=
{\Delta x^{--}}^2(x;x^\prime)
=&(|\eta-\eta^\prime|+{\rm i}\varepsilon)^2-|\mathbf x -\mathbf x^\prime|^2\,.
\end{align}
\end{subequations}
The superscripts $\pm$ are CTP indices that are directly inherited
by the length function through its definition~(\ref{length:fun}).
The superscript $T$ ($\bar T$) indicates (anti-) time ordering,
whereas for the superscript $>$ ($<$), operators evaluated
at the coordinate $x$ ($x^\prime$) appear on the left (right) within
the expression for an expectation value. For more details
on the CTP formalism and its application to quantum fields
in de~Sitter space, see Refs.~\cite{Onemli:2002hr,Prokopec:2003tm,Garbrecht:2011gu}.

The length function (\ref{length:fun}) is useful in order to keep
de~Sitter invariance manifest whenever that is possible. In particular, we
can express the Klein-Gordon equation~(\ref{KG:free})
as
\begin{align}
\label{KG:y}
a^4(x) H^2\left[
-4y\left(1+\frac{y}{4}\right)
\frac{d^2}{d y^2}
-8\left(1+\frac{y}{2}\right)
\frac{d}{dy}
-\frac{m^2_{}}{H^2}
\right]
{\rm i}\Delta^{(0)fg}(y(x;x^\prime))=fg\delta^{fg}{\rm i}\delta^4(\Delta x)\,,
\end{align}
where the exact solution is given by Eq.~(\ref{sol:hypergeo}).
Throughout this work, we are interested the situation where
$m_{}\ll H$, such that sizeable IR fluctuations in the field
$\phi$ occur due to the expansion of the Universe.
In this limit, we can use the approximation
\begin{align}
\label{prop:free:decay}
{\rm i}\Delta^{(0)fg}(y)=\frac{H^2}{4\pi^2}\left[
-\frac{1}{y^{fg}}
+\frac{3 H^2}{2m_{}^2}
\left(-\frac{1}{y^{fg}}\right)^{\frac13\frac{m_{}^2}{H^2}}
+{\cal O}\left(y^{-2}\frac{m^2_{}}{H^2}\right)
\right]
\,,
\end{align}
which follows from the expansion~(\ref{Delta:y:expand}).
(The higher order terms are suppressed by powers of
${m^2_{}}/{H^2}$ and, at large distances, by additional powers of $1/y$.)
The second term in the square brackets is IR enhanced
due to the relation~(\ref{m:lambda:relations:IR:pert}). In order
to calculate the leading IR enhanced corrections to the field
fluctuation, we need to collect
the contribution from the highest power of the IR enhanced factor
at each order in $\lambda$. As we show below, additional IR enhanced
factors result from the space-time integration.

The $\varepsilon$-prescriptions~(\ref{eps:causal}) determine
how to evaluate this solution. In particular,
for time-like ($y>0$) separations, the propagator
${\rm i}\Delta^{(0)}$ acquires an imaginary part.
This can be most conveniently isolated
when making the following approximation
(valid for $m^2_{}\ll H^2$):
\begin{align}
\label{root:y:expand}
\frac{3H^2}{2m^2_{}}(-y)^{-\frac{n}{3}\frac{m^2_{}}{H^2}}
=\frac{3 H^2}{2 m^2_{}}\left(1-{\rm i}\frac{n}{3}\frac{m^2_{}}{H^2}\arg (-y)\right)|y|^{-\frac{n}{3}\frac{m^2_{}}{H^2}}+{\cal O}\left(\frac{m^2_{}}{H^2}\right)\,,
\end{align}
where $n$ is an integer number (arising from powers of the propagator that occur
in Feynman diagrams)
and where
we have suppressed the CTP indices that are responsible for the
infinitesimal phase of $y$. It is useful to note that
\begin{align}
|y|^{-\frac{n}{3}\frac{m^2_{}}{H^2}}
= 1-\frac{n}{3}\frac{m^2_{}}{H^2}\log|y|
+{\cal O}\left(\left(\frac{n}{3}\frac{m^2_{}}{H^2}\log|y|\right)^2\right)\,.
\end{align}
We see that
all basic propagators ${\rm i}\Delta^{(0)++}$, ${\rm i}\Delta^{(0)-+}$, ${\rm i}\Delta^{(0)+-}$
and ${\rm i}\Delta^{(0)--}$ therefore contain an IR-enhanced contribution
$3 H^4/(8\pi^2 m^2_{})$ for
\begin{align}
\label{rel:y:IR}
|y|\ll \exp(3H^2/m^2_{})\,.
\end{align}
For larger values of $|y|$,
the IR-enhanced terms decay following a mild power law. We discuss
below, that this mild decay is however crucial in order to
regulate the integrals that occur within the Schwinger-Dyson
equations. Here, we note in addition that this decay at large distances
is of the same physical origin as the spectral tilt that the inflationary
power spectrum acquires from the $\eta$-parameter~\cite{Liddle:1992wi}.

While the basic propagators all contain IR-enhanced
contributions, within the causal propagators, {\it i.e.} the
retarded and advanced ones
\begin{align}
{\rm i}\Delta^{(0)R,A}(x;x^\prime)={\rm i}\Delta^{(0)T}(x;x^\prime)-{\rm i}\Delta^{(0)<,>}(x;x^\prime)\,,
\end{align}
the IR-enhanced terms cancel.
To obtain a useful representation for the causal propagators,
we note with the help of the approximation~(\ref{root:y:expand}),
that for time-like separations, the propagators
receive non-vanishing imaginary parts through
\begin{subequations}
\begin{align}
{\rm i}\arg (-y^{++}(x;x^\prime))&={\rm i}\pi\vartheta(\Delta x^2)\,,
\\
{\rm i}\arg (-y^{-+}(x;x^\prime))&={\rm i}\pi\vartheta(\Delta x^2){\rm sign}(\Delta x^0)\,,
\\
{\rm i}\arg (-y^{+-}(x;x^\prime))&=-{\rm i}\pi\vartheta(\Delta x^2){\rm sign}(\Delta x^0)\,,
\\
{\rm i}\arg (-y^{--}(x;x^\prime))&=-{\rm i}\pi\vartheta(\Delta x^2)\,,
\end{align}
\end{subequations}
such that we obtain
\begin{subequations}
\label{prop:retav}
\begin{align}
{\rm i}\Delta^{(0)R}(x;x^\prime)&=\frac{H^2}{4\pi^2}
\left[
\frac{1}{y^{<}(x;x^\prime)}
-\frac{1}{y^T(x;x^\prime)}
-{\rm i}\pi\vartheta(\Delta x^2)
\vartheta(\Delta x^0)\left|\frac{1}{y(x;x^\prime)}\right|^{\frac13\frac{m_{}^2}{H^2}}
+\cdots
\right]\,,
\\
{\rm i}\Delta^{(0)A}(x;x^\prime)&=\frac{H^2}{4\pi^2}
\left[
\frac{1}{y^{>}(x;x^\prime)}-\frac{1}{y^T(x;x^\prime)}
-{\rm i}\pi\vartheta(\Delta x^2)
\vartheta(-\Delta x^0)\left|\frac{1}{y(x;x^\prime)}\right|^{\frac13\frac{m_{}^2}{H^2}}
+\cdots
\right]\,.
\end{align}
\end{subequations}

\subsection{Schwinger-Dyson Equations}

The Schwinger-Dyson equations on the CTP are
\begin{align}
\label{SD}
a^4\left(\nabla^2_x-m^2\right){\rm i}\Delta^{fg}(x;x^\prime)
=\delta^{fg}{\rm i}\delta^4(x-x^\prime)-{\rm i}\int d^4w{\rm i}\Pi^{fh}(x;w){\rm i}\Delta^{hg}(w;x^\prime)\,,
\end{align}
where a summation over $h=\pm$ is implied. We have introduced here the full
propagator ${\rm i}\Delta$, to be distinguished
from the free propagator ${\rm i}\Delta^{(0)}$. The self-energy $\Pi$ is derived from
\begin{align}
\Pi^{fg}(x;x^\prime)={\rm i}\frac{\delta\Gamma_2[\Delta]}{\delta\Delta^{gf}(x^\prime;x)}\,,
\end{align}
where $\Gamma_2$ is the two-particle irreducible (2PI) effective action,
that can be computed as $-{\rm i}$ times the sum of all 2PI vacuum diagrams
made up of full propagators.

\begin{figure}[t!]
\begin{center}
\epsfig{file=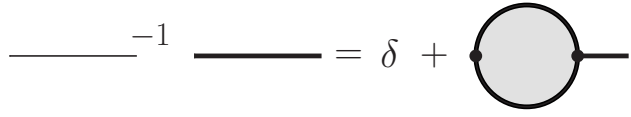,scale=0.6}
\end{center}
\caption{
\label{fig:SD}
Diagrammatic representation of the Schwinger-Dyson equations,
where the thin lines represent free (tree-level) propagators
${\rm i}\Delta^{(0)}$ and
solid lines the full propagators ${\rm i}\Delta$.
The thin line with the superscript
$-1$ denotes the inverse of the free propagator, {\it i.e.} the Klein-Gordon
operator.
The shaded circle represents the sum of all self-energy
diagrams that result from
the functional derivative of the two-particle-irreducible effective action
with respect to the full propagator.
}
\end{figure}

The Schwinger-Dyson equations~(\ref{SD}) can be expressed in terms of diagrams as in
Figure~\ref{fig:SD}. They are exact equations
but in practical calculations one typically aims for approximate solutions.
In this work, we perform a perturbative expansion of the self-energy to
two-loop order. It should therefore
be clear that the propagator ${\rm i}\Delta$ that we obtain below
is only an approximation (which is perturbatively improved
compared to ${\rm i}\Delta^{(0)}$) to the full propagator, even though
we do not introduce an extra symbol for this quantity.

On the CTP, we can take various linear combinations of the Schwinger-Dyson
equations. A particularly useful one is
\begin{subequations}
\label{SDEs}
\begin{align}
\label{Kadanoff-Baym}
a^4(-\nabla^2_x-m^2){\rm i}\Delta^{<,>}(x;x^\prime)=&
-{\rm i}\int d^4w\,{\rm i}\Pi^R(x;w){\rm i}\Delta^{<,>}(w,x^\prime)
\\\notag
&
-{\rm i}\int d^4w\,{\rm i}\Pi^{<,>}(x;w){\rm i}\Delta^A(w,x^\prime)
\,,
\\
a^4(-\nabla^2_x-m^2){\rm i}\Delta^{R,A}(x;x^\prime)=&
{\rm i}\delta^4(x;x^\prime)
-{\rm i}\int d^4w\,{\rm i}\Pi^{R,A}(x;w){\rm i}\Delta^{R,A}(w,x^\prime)
\,,
\end{align}
\end{subequations}
where Eqs.~(\ref{Kadanoff-Baym}) are known as the Kadanoff-Baym equations.

\subsection{Organisation of the Calculation}
\label{sec:organisation}

\begin{figure}[t!]
\begin{center}
\epsfig{file=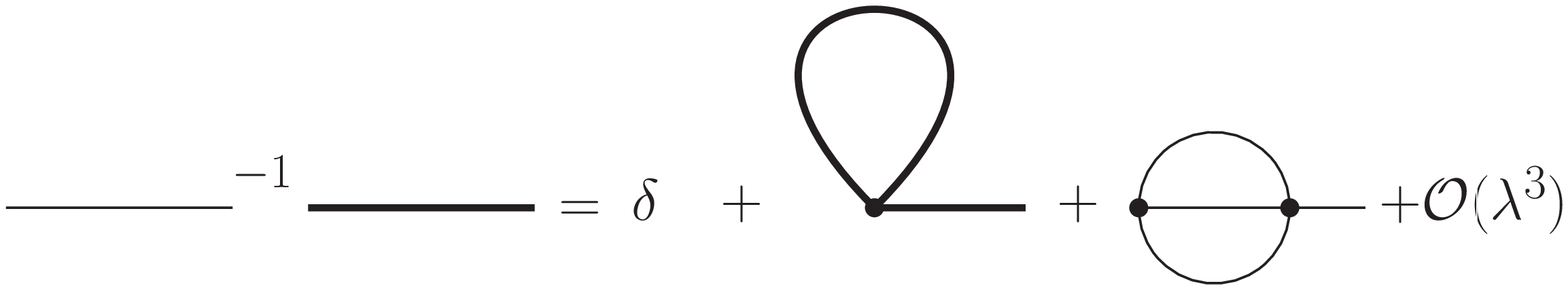,scale=0.6}
\end{center}
\caption{
\label{fig:SD:2}
Diagrammatic representation of the Schwinger-Dyson equations that
are perturbatively truncated at order $\lambda^2$. Again,
thin lines represent free (tree-level) propagators ${\rm i}\Delta^{(0)}$
and solid lines the full propagators ${\rm i}\Delta$. For our present approximation,
that accounts for the leading IR effects to order $\lambda^2$,
it is sufficient to approximate the full propagator
on the right-hand side of these equations by accounting for the seagull-type
mass correction, {\it i.e.} replacing $m^2_{}\to m^2+\delta m^2$ in
Eq.~(\ref{prop:free:decay}).
}
\end{figure}

The main goal of the calculation that is presented here
is to solve the Schwinger-Dyson
equations~(\ref{SDEs})
({\it cf.} also Figure~\ref{fig:SD}) for the propagator ${\rm i}\Delta$.
As these are non-linear integro-differential equations, we aim for
approximate perturbative
solutions that capture the leading IR-effects. For that
purpose, we employ the ansatz that the
loop effects can be approximated by a full propagator that
satisfies the free Klein-Gordon equations with a dynamical mass $m_{\rm dyn}$. This amounts to replacing $m^2\to m^2_{\rm dyn}$ in
Eqs.~(\ref{KG:free},\ref{KG:y},\ref{prop:free:decay}).
Perturbativity is ensured by the relation $\lambda\ll m^4/H^4$.
Then, there are two elementary loop contributions to the self-energy up to
order $\lambda^2$: the seagull diagram that is given in
Figure~\ref{fig:Feyn:diagrams}(A) and the sunset diagram
in Figure~\ref{fig:Feyn:diagrams}(D). We denote the seagull-type
self energy by $\Pi_{\rm sg}$ and the sunset-type by $\Pi_{\rm ss}$.

We now comment on the dynamical mass ansatz and the truncation of the loop
expansion in more detail:
\begin{itemize}
\item
The seagull contribution [Figure~\ref{fig:Feyn:diagrams}(A)] is manifestly local,
and hence it is immediately clear that it takes the effect of a mass correction.
For Euclidean de~Sitter space, it has been demonstrated that the leading IR effects
in massless $\phi^4$ theory, that also include an
infinite number of non-local diagrams,
can be effectively described by a dynamical mass term~\cite{Beneke:2012kn}. The mass
square is then inversely proportional to the fluctuation of $\phi$. For Lorentzian
space, it is shown that local effective equations of motion can be obtained upon
acting on the Schwinger-Dyson equations with the Klein-Gordon
operator~\cite{Garbrecht:2011gu}. However, certain contributions to the effective
equations of motion have been missed in that study, and we show here how to
correctly calculate these for light, perturbative $\phi^4$ theory. The results presented
in this work therefore explicitly show the validity of the dynamical mass ansatz to two-loop order, including in particular the non-local sunset
diagram [Figure~\ref{fig:Feyn:diagrams}(D)].
\item
In the present context, it is useful to review the IR convergence property
of the loop expansion in Euclidean de~Sitter space, because it is rather
straightforward~\cite{Beneke:2012kn}:
Within a Feynman diagram, each propagator contributes an
IR-enhanced factor $\sim H^4/m^2_{}$ and each vertex a factor of
$\lambda$. Adding one vertex to a given diagram, implies two more propagators
and a volume integral yielding a factor $\sim H^{-4}$. Hence, the expansion
parameter can be identified to be $\lambda H^4/m^4_{}$.
For a massless field, it is found that one can replace
$m^2$ with $m^2_{\rm dyn}$, where  $m^2_{\rm dyn}\sim\sqrt \lambda H^2$,
such that diagrams at all loop orders have the same degree of IR
enhancement~\cite{Beneke:2012kn}.
In turn, in presence of a non-vanishing tree-level mass $m$, the loop expansion
is valid provided $\lambda\ll m^4/H^4$, in agreement with the conclusion
from the stochastic approach in Section~\ref{sec:stochastic}.
\item
In Lorentzian space, the seagull diagram has the same degree of IR
enhancement as in Euclidean space~\cite{Riotto:2008mv,Garbrecht:2011gu}.
For the sunset contributions, as
we discuss below, there occur four propagators in the
convolution integrals on the right-hand side of the
Kadanoff-Baym equation~(\ref{Kadanoff-Baym}) (one explicit propagator
and three implicit ones within ${\rm i}\Pi_{\rm ss}$). However, it turns
out that at least one of these propagators is retarded or advanced,
such that it exhibits no IR-enhancement according to Eqs.~(\ref{prop:retav}).
Superficially,
it  may therefore appear that the sunset contribution in Lorentzian space
is less IR-enhanced (by one order in $H^2/m^2_{}$) than in Euclidean
space. As it is shown below, this is however not the case, because the
convolution integral itself contributes an extra factor of
$H^2/m^2_{}$, due to the mild power-law decay of the relevant
contributions to the propagator, {\it cf.} Eq.~(\ref{prop:free:decay}).
Up to two-loop order, the IR convergence properties in Lorentzian
de~Sitter space therefore turn out to effectively
agree with those in Euclidean space, and one
may conjecture that this extends to all orders. For this present calculation,
we therefore take $\lambda\ll m^4/H^4$, such that the perturbation
expansion is valid. This also implies that the number of loops
within a diagram is the same as the order of $\lambda$ in the perturbative
expansion. The relevance of the mild decay of the IR-enhanced terms at large
distances is also emphasised in~\cite{Gautier:2013aoa}.
\item
Now, calculating the leading IR
corrections to the two-point functions up to order
$\lambda^2$ can be done by evaluating the four diagrams
that are given in Figure~\ref{fig:Feyn:diagrams}. Instead, we make use
here of the Schwinger-Dyson equations that are diagrammatically represented
in Figure~\ref{fig:SD}.
When truncated at the perturbative order $\lambda^2$, the Schwinger-Dyson
equations take the form given in Figure~\ref{fig:SD:2}. Note that
the full (bold) propagators are approximated following the
dynamical mass ansatz. As it is indicated in Figure~\ref{fig:SD:2},
we can evaluate
the sunset diagram using the free propagator, because its leading
contribution is of order $\lambda^2$, and additional corrections from
using the full propagator are therefore
of order $\lambda^3$ and higher. Similarly,
for the full propagators that are substituted into the seagull diagram,
we can use an approximation of the dynamical mass-square
$m^2_{\rm dyn}$ that is accurate up
to order $\lambda$, as the leading seagull contribution is already
of order $\lambda$. We denote the seagull diagram with
the full propagator by ${\rm i}\Pi_{\rm SG}$, in order to distinguish it
from the seagull diagram ${\rm i}\Pi_{\rm sg}$ with the free propagator.
Since we choose the parameters such that perturbation theory is valid,
we note the relation
\begin{align}
{\rm i}\Pi_{\rm SG}={\rm i}\Pi_{\rm sg}+{\rm i}\Pi_{\rm ca}+{\cal O}(\lambda^3)\,.
\end{align}
In order to extract the leading IR effects, we only need to keep the
highest powers of the enhancement factor $H^2/m^2$ that occur for each
order in $\lambda$.
\end{itemize}

While the present perturbative calculation does not
make use of the full power of the Schwinger-Dyson approach to
address non-perturbative problems\footnote{In fact, it is
alternatively possible
to calculate the corrections to the correlation function to
${\cal O}(\lambda^2)$ by
simply summing the four diagrams in Figure~\ref{fig:Feyn:diagrams}.},
we yet find it useful because it automatically includes the
Feynman-diagram contributions from Figures~\ref{fig:Feyn:diagrams}(B)
and~\ref{fig:Feyn:diagrams}(C). Moreover, we anticipate that
Schwinger-Dyson equations should be used as well in a future calculation
for massless $\phi^4$ theory in de~Sitter space, that presumably requires
a resummation of diagrams at all loop orders.

\subsection{Solution for a Dynamical Mass Ansatz}

Provided the relation~(\ref{rel:y:IR}) is satisfied,
the propagator~(\ref{prop:free:decay}), which is the solution
to the Klein-Gordon equation~(\ref{KG:y}), can be expressed as
\begin{align}
\label{prop:free:subhor}
{\rm i}\Delta^{(0)fg}(x;x^\prime)=\frac{H^2}{4\pi}\left[-\frac{1}{y^{fg}(x;x^\prime)}-\frac12 \log (-y^{fg}(x;x^\prime)) +\frac{3 H^2}{2 m^2_{}}+\cdots\right]\,.
\end{align}
This is the form of the propagator that is derived in
Ref.~\cite{Prokopec:2003tm} and that is used as well {\it e.g.}
in Refs.~\cite{Garbrecht:2006jm,Garbrecht:2011gu}. It does not exhibit the
mild decay of the approximation~(\ref{prop:free:decay}) for large
separations, when relation~(\ref{rel:y:IR}) is violated.
We show below that the convolution
integrals in the Schwinger-Dyson equations~(\ref{SDEs})
can be analytically evaluated in the coincident approximation
$x\approx x^\prime$, such that as a consequence,
the relation~(\ref{rel:y:IR}) is amply fulfilled and the
approximation~(\ref{prop:free:subhor}) can be used to
infer the fluctuations of the field $\phi$ on horizon scales
and not too far beyond.

Now, we consider a perturbation to
the Klein-Gordon equation~(\ref{KG:y}) by a small correction
$\delta m^2$ to the mass-square as well as by an inhomogeneous constant term
$\mu$:
\begin{align}
\label{KG:y:pert}
&a^4(x) H^2\left[
-4y\left(1+\frac{y}{4}\right)
\frac{d^2}{d y^2}
-8\left(1+\frac{y}{2}\right)
\frac{d}{dy}
-\frac{m^2+\delta m^2}{H^2}
\right]
{\rm i}\Delta^{fg}(y(\Delta x))
\\\notag
=&fg\delta^{fg}\,{\rm i}\delta^4(\Delta x)
+a^4(x)\mu\,.
\end{align}
When identifying the term $\sim \delta m^2$ with the
full seagull correction ${\rm i}\Pi_{\rm SG}$
and the term $\sim \mu$ with the
sunset correction ${\rm i}\Pi_{\rm ss}$ [the precise form
of $\delta m^2$ and $\mu$ is given by
Eqs.~(\ref{mu:specific},\ref{deltam:specific}) below], this equation
corresponds to the Schwinger-Dyson equation that is
truncated at order $\lambda^2$ that is discussed
in Section~\ref{sec:organisation} above and that is represented
by Figure~\ref{fig:SD:2}.

For a solution to Eq.~(\ref{KG:y:pert}) that is
valid on scales where the relation~(\ref{rel:y:IR})
is satisfied, we make the ansatz
\begin{align}
\label{Delta:smally:ansatz}
{\rm i}\Delta(y)=-\frac{H^2}{4\pi^2 y}+A\log y+C+\cdots\,.
\end{align}
The normalisation is imposed by the inhomogeneous $\delta$-function
term, while $A$ and $C$ are coefficients to be determined.
This ansatz is also an approximate solution to the free Klein-Gordon
equation~(\ref{KG:y}) when replacing
${\rm i}\Delta^{(0)}\to {\rm i}\Delta$ and
$m^2\to m^2_{\rm dyn}=\frac{3H^4}{8\pi^2 C}$. This is why it qualifies
as a `dynamical mass' ansatz, even though we find it more convenient
here to approximate the full propagator using the parameter $C$.

Substituting the ansatz~(\ref{Delta:smally:ansatz}) into Eq.~(\ref{KG:y:pert})
and collecting the terms $\propto 1/y$ and the constant terms, we
obtain for the coefficients in the ansatz~(\ref{Delta:smally:ansatz}):
\begin{subequations}
\begin{align}
A=&-\frac{H^2}{8\pi^2}+H^2{\cal O}\left(\frac{m^2+\delta m^2}{H^2}\right)\,,
\\
\label{fluc:inhomc}
C=&-\frac{3H^2}{m^2+\delta m^2}A-\frac{\mu}{m^2+\delta m^2}
\approx\frac{3H^4}{8\pi^2(m^2+\delta m^2)}-\frac{\mu}{m^2}
\\\notag
\approx&\frac{3H^4}{8\pi^2 m^2}-\frac{3H^4 \delta m^2}{8\pi^2 m^4}+\frac{3H^4 \delta m^4}{8\pi^2 m^6}-\frac{\mu}{m^2}\,.
\end{align}
\end{subequations}
The last two approximations are valid to order $\lambda^2$,
when we identify $\delta m^2\sim{\rm i}\Pi_{\rm SG}$
and $\mu\sim{\rm i}\Pi_{\rm ss}$. (These identifications are made
more explict and quantitative in Section~\ref{massive:FT}.)
Note that we need to account for the
dependence of $\delta m^2$ on $C$, in order to obtain a self-consistent
solution for $C$, that is valid up to order $\lambda^2$.
In fact, we
can then interpret within the last expression of Eq.~(\ref{fluc:inhomc})
the second term as the sum of the seagull and the cactus diagram
in Figures~\ref{fig:Feyn:diagrams}(A) and~\ref{fig:Feyn:diagrams}(C) (provided $\Pi_{\rm SG}$ and hence $\delta m^2$ are accurate
to order $\lambda^2$), the third term as the diagram
in Figure~\ref{fig:Feyn:diagrams}(B) and the fourth term as
the sunset diagram, Figure~\ref{fig:Feyn:diagrams}(D). This shows that
indeed, the Schwinger-Dyson approach is equivalent here to a
straightforward perturbative calculation.
We also observe that at leading order in $H^2/m^2$,
a mass-square perturbation $\delta m^2$
takes the same effect as a constant inhomogeneous term
$8\pi^2\mu m^2/(3H^4)$.





\subsection{Evaluation of the Self-Energies and Solution to the Schwinger-Dyson equations}
\label{massive:FT}

Now, we show that the leading IR contributions to the self-energy terms on the right
hand side of the Kadanoff-Baym equations~(\ref{Kadanoff-Baym})
can indeed be effectively accounted for by the terms $\delta m^2$ and
$\mu$ as in the perturbed
the Klein-Gordon equation~(\ref{KG:y:pert}). We moreover
evaluate these corrections and compare the result for the full propagator
to the stochastic fluctuations.

The sunset contribution [Figure~\ref{fig:Feyn:diagrams}(D)]
to the self-energy is given by
\begin{align}
{\rm i}\Pi_{\rm ss}^{fg}(x;x^\prime)=\frac{\lambda^2}{6}a^4(x)\left({\rm i}\Delta^{(0)fg}(x;x^\prime)\right)^3a^4(x^\prime)\,.
\end{align}
Using this expression, we evaluate the first of the convolution
integrals in Eq.~(\ref{Kadanoff-Baym}) for $x=x^\prime$:
\begin{align}
\label{convul:1}
&-{\rm i}\int d^4 w {\rm i}\Pi_{\rm ss}^{<,>}(x;w){\rm i}\Delta^{(0)A}(w;x)
\\\notag
=&-{\rm i}\int d^4 w \frac{\lambda^2}{6}
a^4(x)\left({\rm i}\Delta^{(0)<,>}(x;w)\right)^3a^4(w){\rm i}\Delta^{(0)A}(w;x)
\\\notag
\approx&-\frac{9\lambda^2 H^{14}}{2^{12}\pi^8 m_{}^6}a^4(x)
\int d^4 w a^4(w)\left|y(x;w)\right|^{-\frac43\frac{m^2_{}}{H^2}}
\pi\vartheta((w^0-x^0)^2-(\mathbf w-\mathbf x)^2)\vartheta(w^0-x^0)
\\\notag
\approx&-a^4(x)\frac{9\lambda^2 H^{12}}{2^{12}\pi^6m^8_{}}\,.
\end{align}
For the first approximation, we have extracted the leading IR contributions
that are enhanced by factors of $H^2/m^2$.
The details for the calculation of the last integral and the
approximations made are presented in Appendix~\ref{appendix:convolution}.
A similar
calculation yields an approximation for the second convolution
integral in Eq.~(\ref{Kadanoff-Baym}):
\begin{align}
\label{convul:2}
-{\rm i}\int d^4 w {\rm i}\Pi_{\rm ss}^{R}(x;w){\rm i}\Delta^{(0)<,>}(w;x)
\approx&-a^4(x)\frac{27\lambda^2 H^{12}}{2^{12}\pi^6m^8_{}}\,.
\end{align}
We note here that while only three explicitly IR-enhanced factors
$H^2/m^2_{}$ occur within the integrands of the convolution
integrals,
as the retarded and advanced propagators~(\ref{prop:retav}) do not contain such
terms, the results of the integrals are nonetheless of fourth order
in $H^2/m^2_{}$. This is because the integration
yields the missing factor due to the mild decay of the
propagators at large distances, which can be seen from
Eq.~(\ref{prop:free:decay}).
By comparing
Eq.~(\ref{KG:y:pert}) with Eq.~(\ref{Kadanoff-Baym}), we identify
\begin{align}
\label{mu:specific}
\mu=-\frac{9\lambda^2H^{12}}{2^{10}\pi^6 m^8}+{\cal O}(\lambda^3)\,.
\end{align}
Note that $-\mu$ agrees with the corresponding result
$\langle\phi^2\rangle_{\rm ss}$ from the sunset
diagram in the stochastic approach Eq.~(\ref{ss1}). 

We also quote the the leading IR results for the seagull-type self-energies
and convolutions,
${\rm i}\Pi_{\rm SG}^{<,>}(x;w)=0$
and
\begin{align}
-{\rm i}\int d^4 w {\rm i}\Pi_{\rm SG}^{R}(x;w){\rm i}\Delta^{<,>}(w;x^\prime)
=&a^4(x) \frac{\lambda}{2} C {\rm i}\Delta(x;x^\prime)=
a^4(x)\delta m^2 {\rm i}\Delta(x;x^\prime)\,,
\end{align}
such that
\begin{align}
\label{deltam:specific}
\delta m^2=\frac{\lambda}{2} C\,.
\end{align}
We note that the term $-3H^4\delta m^2/(8\pi^2 m^4)$
in Eq.~(\ref{fluc:inhomc})
can be identified with the sum of the seagull and the cactus diagram,
Figures~\ref{fig:Feyn:diagrams}(A) and~\ref{fig:Feyn:diagrams}(C). Furthermore,
the result agrees with the corresponding diagrammatic contributions
in the stochastic approach,
$\langle \phi^2\rangle_{\rm sg}+\langle \phi^2\rangle_{\rm ca}$,
see Eqs.~(\ref{sg},\ref{ca}). Similarly, the term
$3H^4\delta m^4/(8\pi^2 m^6)$ can be identified with the diagram
in Figure~\ref{fig:Feyn:diagrams}(C), and it agrees with the stochastic result,
$\langle \phi^2\rangle_{\rm sgsg}$,
Eq.~(\ref{sgsg1}). Together with the fact that
$-\mu^2=\langle \phi^2\rangle_{\rm ss}$, this implies that to
${\cal O}(\lambda^2)$, the
individual Feynman diagrams in the CTP formalism (after summing over CTP indices on internal vertices) evaluate
to the same results as the corresponding diagrams in the stochastic approach.
Of course, it would be interesting to generalize this statement
beyond the two-loop order.

Now, we substitute Eqs.~(\ref{mu:specific})
and~(\ref{deltam:specific}) into Eq.~(\ref{fluc:inhomc}) and solve for the
coefficient $C$, such that we obtain
\begin{align}
\label{result:C}
C=&\frac{3H^4}{8 \pi^2 m^2}
-\frac{9 \lambda H^8}{128 \pi^4 m^6}
+\frac{9 \lambda^2 H^{12}}{256 \pi^6 m^{10}}+\cdots\,,
\end{align}
which is in agreement with the result~(\ref{lambda-series1})
derived by stochastic methods.
Note that we may also infer a dynamical mass through
\begin{align}
m^2_{\rm dyn}=&
\frac{3 H^4}{8\pi^2 C}
=m^2+\frac{3\lambda H^4}{16\pi^2 m^2}-\frac{15\lambda^2 H^8}{256\pi^4 m^6}+{\cal O}(\lambda^3)\,.
\end{align}

\section{Generalisation to $D$ Space-Time Dimensions}
\label{sec:Ddim}

We now explain how to generalise above calculation to $D$ space-time dimensions
and present the result for the field fluctuations. In $D$ dimensional de~Sitter space, the action is given by
\begin{align}
S=\int d^D x {\cal L}
\end{align}
and the Klein-Gordon equation is:
\begin{align}
\label{KG:y:D}
a^D(x)H^2\left[-4y\left(1+\frac{y}{4}\right)\frac{d^2}{dy^2}-2D\left(1+\frac{y}{2}\right)\frac{d}{dy}-\frac{m^2_{}}{H^2}\right]{\rm i}\Delta^{(0)fg}(y)
=fg\delta^{fg}\delta^D(\Delta x)\,.
\end{align}
We can also generalise Eq.~(\ref{KG:y:pert}) to $D$ dimensions by replacing
in Eq.~(\ref{KG:y:D}) ${\rm i}\Delta^{(0)}\to{\rm i}\Delta$,
$m^2_{}\to m^2+\delta m^2$ and adding the term $a^D(x) \mu$
to the right-hand side.
For the solution, we make the ansatz
\begin{align}
\label{ansatz:Delta:Ddim}
{\rm i}\Delta(y)=-\frac{2}{(D-1) K H^2 y}+A\log y +C+\cdots\,,
\end{align}
where the normalisation is imposed again by the inhomogeneous
$\delta$-function term in Eq.~(\ref{KG:y:D}), and where we define
\begin{align}
\label{def:K}
K=\frac{2\pi^{\frac{D+1}{2}}}{\Gamma\left(\frac{D+1}{2}\right)H^D}\,.
\end{align}
Substituting the ansatz~(\ref{ansatz:Delta:Ddim}) into
Eq.~(\ref{KG:y:D}) and comparing the coefficients then yields
\begin{align}
A=-\frac{1}{(D-1)K H^2}+\cdots
\end{align}
and
\begin{align}
\label{fluc:inhomC:D}
C=-\frac{(D-1)H^2}{m^2+\delta m^2}A-\frac{\mu}{m^2+\delta m^2}
\approx \frac{1}{K(m^2+\delta m^2)}-\frac{\mu}{m^2+\delta m^2}\,.
\end{align}

It remains to determine the $D$-dimensional expressions
for $\delta m^2$ and $\mu$.
The expression~(\ref{deltam:specific}) for $\delta m^2$ is valid in
$D$ dimensions as well. For the convolution integrals,
we obtain
\begin{subequations}
\label{convul:Ddim}
\begin{align}
-{\rm i}\int d^D w{\rm i}\Pi^{<,>}_{\rm ss}(x;w){\rm i}\Delta^{(0)A}(w;x)
=&-{\rm i}\int d^D w \frac{\lambda^2}{6}a^D(x){\rm i}\left(\Delta^{(0)<,>}(x;w)\right)^3a^D(w)
\\\notag
\times&{\rm i}\Delta^{(0)A}(w;x)
=-a^D(\eta)\frac{\lambda^2}{24 K^3 m^8}\,,
\\
-{\rm i}\int d^D w{\rm i}\Pi^{R}_{\rm ss}(x;w){\rm i}\Delta^{(0)<,>}(w;x)=&-a^D(\eta)\frac{\lambda^2}{8 K^3 m^8}\,.
\end{align}
\end{subequations}
We can hence identify
\begin{align}
\mu=-\frac{\lambda^2}{6K^3m^8}\,.
\end{align}
Substituting this result and the expression~(\ref{deltam:specific})
for $\delta m^2$ into Eq.~(\ref{fluc:inhomC:D}), we obtain
\begin{align}
\label{result:C:Ddim}
C=\frac{1}{K m^2}-\frac{\lambda}{2K^2 m^6}+\frac{2\lambda^2}{3 K^3 m^{10}}+\cdots\,.
\end{align}
Note that within the term of ${\cal O}(\lambda^2)$,
four factors of $K^{-1}$ originate from the fourth power of
the propagator given in Eq.~(\ref{Delta:y:expand:D}) while
a factor of $K$ results from the convolution integral. The fact
that up to order $\lambda^2$ (and perhaps beyond) the dependence of
the term of order $\lambda^n$ on
the space-time dimension is proportional
to $K^{-n-1}$ is therefore a non-trivial result of the QFT
calculation. In contrast, this feature emerges from the stochastic approach
in a rather obvious manner.
We can also turn the result for $C$ around in order to obtain a dynamical mass
\begin{align}
\label{mdyn:Ddim}
m^2_{\rm dyn}=\frac{1}{K C}=m^2+\frac{\lambda}{2 K m^2}-\frac{5 \lambda^2}{12 K^2 m^6}+\cdots\,.
\end{align}
Note that the dynamical mass is inferred here from the field
fluctuations, whereas the effective infrared mass
in Ref.~\cite{Gautier:2013aoa} governs the decay of the two-point function
in the far infrared. The two quantities are therefore not directly comparable
and do in fact differ~\footnote{We thank F.~Gautier and J.~Serreau for pointing
this out.}.
As for the details of the
calculation, in the present work, we remain  in position space and aim to obtain
the intermediate and final results in a manifestly
de~Sitter invariant manner, whenever that is
possible. A particular reference frame [through the conformal coordinates in
Eq.~(\ref{dS:metric})] has to be chosen however
in order to perform the convolution integrals as
described in Appendix~\ref{appendix:convolution}. At that stage, it appears unavoidable
to give up the manifest de~Sitter invariance.
In contrast, in
Ref.~\cite{Gautier:2013aoa}, a mixed representation for the two-point functions is chosen, where the spatial coordinates are expressed in momentum space, and manifest
de~Sitter invariance is lost throughout the calculation. Of course, the final
results should preserve de~Sitter invariance in both approaches.

Again, we compare the field fluctuation~(\ref{result:C:Ddim})
with the corresponding result from the stochastic approach.
The 4-dimensional derivation for the stochastic
fluctuations from
Refs.~\cite{Starobinsky:1986fx,Starobinsky:1994bd} is generalised
to $D$ dimensions
in Ref.~\cite{Beneke:2012kn}, where it is shown that the appropriate probability
distribution is
\begin{align}
\varrho(\phi)={\cal N}{\rm e}^{-K V(\phi)}\,.
\end{align}
Using this within Eqs.~(\ref{PDF:normalisation})
and~(\ref{phi:expectationvalue}), we obtain that
\begin{align}\label{stoch:D}
\langle \phi^2 \rangle=\frac{1}{K m^2}-\frac{\lambda}{2K^2 m^6}+\frac{2\lambda^2}{3 K^3 m^{10}}+\cdots\,,
\end{align}
in agreement with the result~(\ref{result:C:Ddim}) derived by
QFT methods.

\section{Conclusions}

The main results of this paper are the scalar field
fluctuations~(\ref{result:C}), (\ref{result:C:Ddim})
obtained by QFT methods and the fact that these agree with the corresponding
quantities derived by the stochastic method, eqs.~(\ref{lambda-series1}) and (\ref{stoch:D}). While it is well known that
the two methods yield the same results
for the leading IR-enhanced corrections from the local
seagull diagram [Figure~\ref{fig:Feyn:diagrams}(A)], which is of relevance in the large
$N$ limit~\cite{Riotto:2008mv,Serreau:2011fu}, this is to our knowledge
the first result where this agreement is generalised to a
situation where a non-local self-energy diagram, {\it i.e.} the seagull
graph from Figure~\ref{fig:Feyn:diagrams}(C), is involved. In fact,
when breaking down the perturbative corrections to the individual Feynman
diagrams in Figure~\ref{fig:Feyn:diagrams}, all corresponding graphs obtained
by the two methods are found to agree.

While there appears to be a close relation between the expressions that
one obtains when calculating the fluctuations in Euclidean de~Sitter
space to those from the stochastic approach~\cite{Beneke:2012kn},
in that the IR enhancement
results entirely from powers of Gaussian two-point functions,
accounting for the IR enhancement in Lorentzian de~Sitter space is
a bit less straightforward: Although there is no leading order IR enhancement from
the retarded and advanced propagators~(\ref{prop:retav}), the lack of IR
power is enhanced by a space-time integral that is regulated
only by the mild decay of the two-point function at large
separations~(\ref{prop:free:decay}). The importance of the mild decay
behaviour is also emphasised in Ref.~\cite{Gautier:2013aoa}.
However, when comparing
the result obtained there for the dynamical mass
with Eq.~(\ref{mdyn:Ddim}),
there appears to be disagreement at ${\cal O}(\lambda^2)$.
In the CTP formalism, there are several different terms arising
from a single Feynman diagram, due to the summation over the CTP
indices. As a consequence, there are two different terms on the
right-hand side of the Kadanoff-Baym equations~(\ref{Kadanoff-Baym}).
For the sunset diagram [Figure~\ref{fig:Feyn:diagrams}(D)], we
need to evaluate these here separately and we find that both
terms are relevant at the same order of IR enhancement, but they
come with
different coefficients.

To this end, the evaluation of particular Feynman diagrams therefore appears
more involved in Lorentzian de~Sitter space than in its Euclidean counterpart,
which is not surprising, as the same can also be stated about
field theory in flat space-times. Nonetheless, the agreement at the
level of Feynman diagrams up to two-loop order suggests that this observation
may be generalised to arbitrary loop order by extending the methods
that are presented here. In Ref.~\cite{Beneke:2012kn},
it is already shown how to resum the individual Euclidean Feynman diagrams
to match the result from the stochastic approach. Therefore, it appears
that the step that is yet missing in order to explicitly solve for
the leading order IR behaviour of the vacuum state of massless
$\phi^4$ theory in de~Sitter space may be to demonstrate that the agreement
between QFT and stochastic Feynman diagrams found here extends to all loop
orders. If the agreement between the QFT and the stochastic methods at
the diagrammatic level up to two-loop order is not accidental, then it is
remarkable that the stochastic formalism vastly simplifies the way the
leading IR effects are accounted for. It would be interesting to perform
a derivation of the stochastic expressions for obtaining correlation functions that relies on the CTP formalism and that
does not appeal to the classical behaviour
of the superhorizon modes.
This may also point to a simplification of the evaluation of the Feynman diagrams in the CTP formalism in de~Sitter
background.
Work in that direction has been performed
in Ref.~\cite{Rigopoulos:2013exa}, and we will further
address this point in the future~\cite{GRZ}.

While in the  work~\cite{Rajaraman:2010xd,Beneke:2012kn} the problem
of massless $\phi^4$ theory in Euclidean space is resolved at leading
IR order, and since arguments
in favour of an analytic continuation of correlation functions
to Lorentzian space exist~\cite{Higuchi:2010xt}, the efforts to
obtain a direct solution to the problem in Lorentzian space are justified,
because they lead to insights into the dynamical behaviour of
the model~\footnote{Beyond the leading IR order, it is reported that perturbation theory in Euclidean de~Sitter space indicates correlation functions that grow in temporal
separations~\cite{Hollands:2011we}.
For this reason, and also for obtaining corrections beyond the
stochastic approximation, it would be interesting to calculate sub-leading contributions in the IR expansion of the Lorentzian correlation functions, using the methods that are employed in the present work.}. This is of particular relevance for Cosmology, where the
initial inflationary stage, that is often presumed to set the initial
conditions for the subsequent evolution of the Universe, deviates from
de~Sitter space in that the Hubble rate is time-dependent, and that
the approximate de~Sitter epoch comes to an end and perhaps also
begins at a finite time. These time dependences cannot be
directly accounted for in the Euclidean approach.
Nonetheless, they have important consequences
for the observed perturbation spectrum and perhaps lead to a significant backreaction~\cite{Tsamis:1993ub,Onemli:2002hr,Onemli:2004mb,Tsamis:1996qm,Tsamis:1996qq}.
It therefore remains an important task to find the
correct solutions for massless $\phi^4$ theory and eventually also Gravitation
in Lorentzian de~Sitter space.

\subsubsection*{Acknowledgements}
\noindent
We would like to thank M.~Beneke, P.~Moch F.~Gautier and J.~Serreau
for discussions and for
useful comments on the manuscript.
This work is supported by the Gottfried Wilhelm Leibniz programme
of the Deutsche Forschungsgemeinschaft
and by the DFG cluster of excellence ‘Origin and Structure of the Universe’.

\begin{appendix}
\numberwithin{equation}{section}

\section{Long-Distance Behaviour of the Propagator}

The well-known solution to Eq.~(\ref{KG:y:D})
[and Eq.~(\ref{KG:y}) for $D=4$] is given in terms of
a hypergeometric function:
\begin{align}
\label{sol:hypergeo}
{\rm i}\Delta^{(0)}(y)=
\frac{\Gamma\left(\frac{D-1}{2}+\nu\right)\Gamma\left(\frac{D-1}{2}-\nu\right)}{(4\pi)^{\frac{D}{2}}\Gamma\left(\frac{D}{2}\right)}
H^{D-2}{_2F_1}\left(\frac32+\nu,\frac32-\nu;2;1+\frac y4\right)\,,
\end{align}
where
\begin{align}
\nu=\sqrt{\left(\frac{D-1}{2}\right)^2-\frac{m^2_{}}{H^2}}\,.
\end{align}
In order to extract the leading behaviour for large $y$, we make use
of the transformation formula
\begin{align}
&{_2F_1}\left(\frac{D-1}{2}+\nu,\frac{D-1}{2}-\nu;\frac{D}{2};1+\frac y4\right)
\\\notag
=&
\frac{\left(-\frac y4\right)^{-\frac{D-1}{2}-\nu}\Gamma\left(\frac{D}{2}\right)\Gamma(-2\nu)}{\Gamma\left(\frac{D-1}{2}-\nu\right)\Gamma\left(\frac12-\nu\right)}
{_2F_1}\left(\frac{D-1}{2}+\nu,\frac12+\nu;1+2\nu;-\frac4y\right)
\\\notag
+&
\frac{\left(-\frac y4\right)^{-\frac{D-1}{2}+\nu}\Gamma\left(\frac{D}{2}\right)\Gamma(2\nu)}{\Gamma\left(\frac{D-1}{2}+\nu\right)\Gamma\left(\frac12+\nu\right)}
{_2F_1}\left(\frac{D-1}{2}-\nu,\frac12-\nu;1-2\nu;-\frac4y\right)\,.
\end{align}
The desired approximation then results from the defining series
expansion
\begin{align}
{_2F_1}(\alpha,\beta;\gamma;z)=\frac{\Gamma(\gamma)}{\Gamma(\alpha)\Gamma(\beta)}
\sum\limits_{n=0}^{\infty}\frac{\Gamma(\alpha+n)\Gamma(\beta+n)}{\Gamma(\gamma+n)}\frac{z^n}{n!}\,,
\end{align}
such that we obtain for $D=4$
\begin{align}
\label{Delta:y:expand}
{\rm i}\Delta^{(0)}(y)=&
H^2\left(\frac{1}{16\pi^2}-\frac{1}{24\pi^2}\frac{m^2_{}}{H^2}+{\cal O}\left(\frac{m^4_{}}{H^2}\right)\right)\left(-\frac y4\right)^{-\frac52+\nu}
\\\notag
+&H^2\left(\frac{3H^2}{8\pi^2m^2_{}}-\frac{7}{24\pi^2}+{\cal O}\left(\frac{m^2_{}}{H^2}\right)\right)\left(-\frac y4\right)^{-\frac32+\nu}
\\\notag
+&H^2{\cal O}\left(\frac{m^2_{}}{H^2}\right)\left(-\frac y4\right)^{-\frac72+\nu}+\cdots
\,.
\end{align}
When keeping $D$ general, the expressions for the higher-order terms, which are
not relevant in the present context, are
somewhat lengthy. We therefore only present the leading behaviour, which is
\begin{align}
\label{Delta:y:expand:D}
{\rm i}\Delta^{(0)}(y)=\frac{1}{2(D-1)KH^2}\left(-\frac{y}{4}\right)^{-1-\frac{m^2_{}}{(D-1)H^2}}
+\frac{1}{m^2_{} K}\left(-\frac{y}{4}\right)^{-\frac{m^2_{}}{(D-1)H^2}}
+\cdots\,,
\end{align}
and where $K$ is defined in Eq.~(\ref{def:K}).

\section{Convolution integral in the Kadanoff-Baym Equation}
\label{appendix:convolution}


The integrations in Eqs.~(\ref{convul:1},\ref{convul:2}) can be approximately performed
as follows:
\begin{align}
{\cal I}(x)=&
a^D(x)
\int d^D x^\prime a^D(x^\prime)\left|y(x;x^\prime)\right|^{-\frac{n}{D-1}\frac{m^2_{}}{H^2}}
\vartheta((\eta-\eta^\prime)^2-(\mathbf x-\mathbf x^\prime)^2)\vartheta(\eta^\prime-\eta)
\\\notag
=&
\frac{1}{H^D\eta^D}
\int\limits_{-\infty}^\eta d\eta^\prime\frac{1}{H^D{\eta^\prime}^D}
\int d^{D-1} x^\prime \vartheta((\eta-\eta^\prime)^2-(\mathbf x-\mathbf x^\prime)^2)
\\\notag
&\hskip5cm \times
\left(\frac{\eta\eta^\prime}{\left[(\eta-\eta^\prime)^2-(\mathbf x-\mathbf x^\prime)^2\right]}\right)^{\frac{n}{D-1}\frac{m^2_{}}{H^2}}
\\\notag
=&
\frac{1}{H^D\eta^D}\int\limits_{-\infty}^\eta d\eta^\prime \frac{2\pi^{\frac{D-1}{2}}(\eta \eta^\prime)^{\frac{n}{D-1}\frac{m^2_{}}{H^2}}}{H^D{\eta^\prime}^D\Gamma\left(\frac{D-1}{2}\right)}
(\eta-\eta^\prime)^{D-1-\frac{2n}{D-1}\frac{m^2_{}}{H^2}}
\frac{\Gamma\left(\frac{D-1}{2}\right)\Gamma\left(1-\frac{n}{D-1}\frac{m^2_{}}{H^2}\right)}{2\Gamma\left(\frac{D+1}{2}-\frac{n}{D-1}\frac{m^2_{}}{H^2}\right)}
\\\notag
\approx&
\frac{1}{H^{2D}\eta^D}\frac{\pi^{\frac{D-1}{2}}}{\Gamma\left(\frac{D+1}{2}\right)}
\int\limits_{-\infty}^\eta d\eta^\prime (-\eta^{\prime})^{-1-\frac{n}{D-1}\frac{m^2_{}}{H^2}}
=\frac{\pi^{\frac{D-1}{2}}}{H^{2D}\eta^D\Gamma\left(\frac{D+1}{2}\right)}
\left[
\frac{(-\eta^\prime)^{-\frac{n}{D-1}\frac{m^2_{}}{H^2}}}{\frac{n}{D-1}\frac{m^2_{}}{H^2}}
\right]^{\eta^\prime=\eta}_{\eta^\prime\to-\infty}
\\\notag
\approx&a^D(\eta)\frac{(D-1)\pi^{\frac{D-1}{2}}}{n \Gamma\left(\frac{D+1}{2}\right)H^{D-2} m^2_{}}\,.
\end{align}
We also make use of this result in order to obtain Eqs.~(\ref{convul:Ddim}).

\end{appendix}

\end{document}